\documentclass[letter,oldversion] {aa} 
\usepackage{amsmath}
\usepackage{graphicx}
\usepackage{color}
\usepackage{txfonts}
\usepackage[authoryear]{natbib}

\begin{document}

\title{Inversion of the decay of the cyclotron line energy in Her X-1}

\author{R.~Staubert\inst{1}, D.~Klochkov\inst{1}, F.~F\"urst\inst{2}, J.~Wilms\inst{3},  
R.E.~Rothschild\inst{4}, F.~Harrison\inst{5}}

\offprints{staubert@astro.uni-tuebingen.de}

\institute{
        Institut f\"ur Astronomie und Astrophysik, Universit\"at T\"ubingen,
        Sand 1, 72076 T\"ubingen, Germany
\and
        European Space Agency - European Space Astronomy Center (ESA-ESAC), Operations Dpt.,
        Camino Bajo del Castillo, s/n., Urb.\ Villafranca del Castillo, 28692 Villanueva de la Canada, 
        Madrid, Spain 
\and
        Dr.\ Remeis Sternwarte, Astronomisches Institut der
        Universit\"at Erlangen-N\"urnberg, Sternwartstr.~7, 96049 Bamberg, Germany
\and
        Center for Astrophysics and Space Sciences, University of
        California at San Diego, La Jolla, CA 92093-0424, USA
\and
        Cahill Center for Astronomy and Astrophysics, California Institute of Technology, 
        Pasadena, CA 91125, USA 
}

\date{submitted: 11/09/2017, accepted: 09/10/2017 }
\authorrunning{Staubert et al.}
\titlerunning{Inversion of cyclotron line energy in Her X-1}

\abstract
   {Recent observations of Her~X-1 with \textsl{NuSTAR} and \textsl{INTEGRAL} in 2016
   have provided evidence that the 20-year  decay of the cyclotron line energy found between
   1996 and 2015 has ended and that an inversion with a new increase, possibly similar to the one 
   observed around 1990--1993, has started. We consider this a strong motivation for further 
   observations and for enhanced efforts to significantly improve our theoretical understanding of 
   the accretion process in binary X-ray pulsars. We speculate about the physics behind the
   long-term decay and its inversion, a possible cyclic behavior, and correlations with other
   variable observables.
   }

   
   \maketitle
%

\section{Introduction}

The X-ray spectrum of the accreting binary pulsar Her~X-1 is
characterized by a power law continuum with exponential cutoff and an
apparent line-like feature, which was discovered in 1975
\citep{Truemper_etal78}. This feature is now generally accepted as an
absorption feature around 40\,keV due to resonant scattering of
photons off electrons on quantized energy levels (Landau levels) in
the teragauss magnetic field at the polar cap of the neutron star. The
feature is therefore often referred to as a cyclotron resonant
scattering feature (CRSF). The energy spacing between the Landau
levels is given by $E_\mathrm{cyc} = \hbar eB/m_{\rm e}c =
11.6\,\text{keV}\,B_{12}$, where $B_{12}$ is the strength of the
magnetic field in units of $10^{12}\,\text{Gauss}$, providing a direct
method of measuring the magnetic field strength at the site of the
emission of the X-ray spectrum. The observed line energy is subject to
gravitational redshift, $z$, such that the magnetic field may be
estimated by $B_{12} = (1+z) E_\text{obs}/11.6\,{\rm keV}$. The
discovery of the cyclotron feature in the spectrum of Her X-1 provided
the first ever direct measurement of the magnetic field strength of
a neutron star, in the sense that no other model assumptions are
needed. Originally considered an exception, cyclotron features are now
known to be rather common in accreting X-ray pulsars;  $\sim 35$
binary pulsars are now known to be cyclotron line sources. In several
objects, multiple lines (up to four harmonics) have been detected
(for reviews, see: \citealt{Coburn_etal02,Staubert_03,Heindl_etal04,
Terada_etal07,Wilms_12,CaballeroWilms_12,RevnivtsevMereghetti_16}).

\begin{table}
\caption{Recent cyclotron line energy measurements in Her~X-1 by
  \textsl{NuSTAR} and \textsl{INTEGRAL}. Uncertainties are at the 68\% level.
  The maximum flux in the respective 35-day cycle is given in units of ASM-cts/s,
  referring to the All Sky Monitor of \textsl{RXTE}. The flux was actually measured
  by \textsl{Swift}/BAT and converted according to
  $\mbox{(2--10\,keV)} ASM-cts\,\mathrm{s}^{-1}=93.0\times \mbox{(15--50\,keV)} BAT-cts\,\mathrm{cm}^{-2}\,\mathrm{s}^{-1}$
  \citep{Staubert_etal16}. 35-day cycle numbering is according to \citet{Staubert_etal83}.}
\vspace{-3mm}
\begin{center}
\begin{tabular}{lllll}
\hline\noalign{\smallskip}
Observation       & 35\,d & Center   & Obs. Line       & max. Flux \\
month/year         & cycle & MJD      & Energy [keV]  & [ASM-cts/s] \\
\hline\noalign{\smallskip}
\textsl{NuSTAR} \\       
21 Aug 2016    & 468 & 57621 & $37.29\pm0.14$  & $6.09\pm0.18$ \\
\textsl{INTEGRAL} \\   
03--04 Apr 2016 & 464 & 57481 & $37.31\pm0.81$  & $6.16\pm0.36$ \\
19--21 Aug 2016 & 468 & 57621 & $38.16\pm0.70$  & $6.09\pm0.18$ \\
\noalign{\smallskip}\hline
\end{tabular}\end{center}
\label{tab:obs}
\end{table}

The Her X-1/HZ Her binary system shows strong variability on very
different  timescales: there is the 1.24\,s spin period of the
neutron star, the 1.7\,d binary period, the 35\,d flux modulation,
and the 1.65\,d period of the pre-eclipse dips. 
The 35\,d \textsl{On-Off} variation can be understood as being due to the precession
of a warped accretion disk. Due to the high inclination 
of the binary ($i>80^\circ$) we see the disk nearly edge-on. The precessing warped
disk therefore covers the central X-ray source during a substantial
portion of the 35\,d period. Furthermore, a hot X-ray heated accretion
disk corona reduces the X-ray signal (energy independently) by Compton
scattering whenever it intercepts our line of sight to the neutron
star. As a result, the X-ray source is covered twice during a 35\,d
cycle. A further modulation appears through the so-called
\textsl{Anomalous Lows} (ALs) that reduce the X-ray flux to
unobservable levels for time periods ranging from days to years (see
Table~\ref{tab:AL} for a summary of  information about the ALs to date). The ALs are probably caused by a low inclination of
the accretion disk, possibly combined with a thickening of the inner
part of the accretion disk, thereby shading the X-ray emitting regions
at the poles of the neutron star. The ALs tend to appear on a
quasi-period of $\sim$5.5\,yr if we assume that two out of seven
ALs were actually not realized (at least not observed).

\begin{figure*}
\label{fig:Ecyc}
\resizebox{\hsize}{!}{\includegraphics[angle=90]{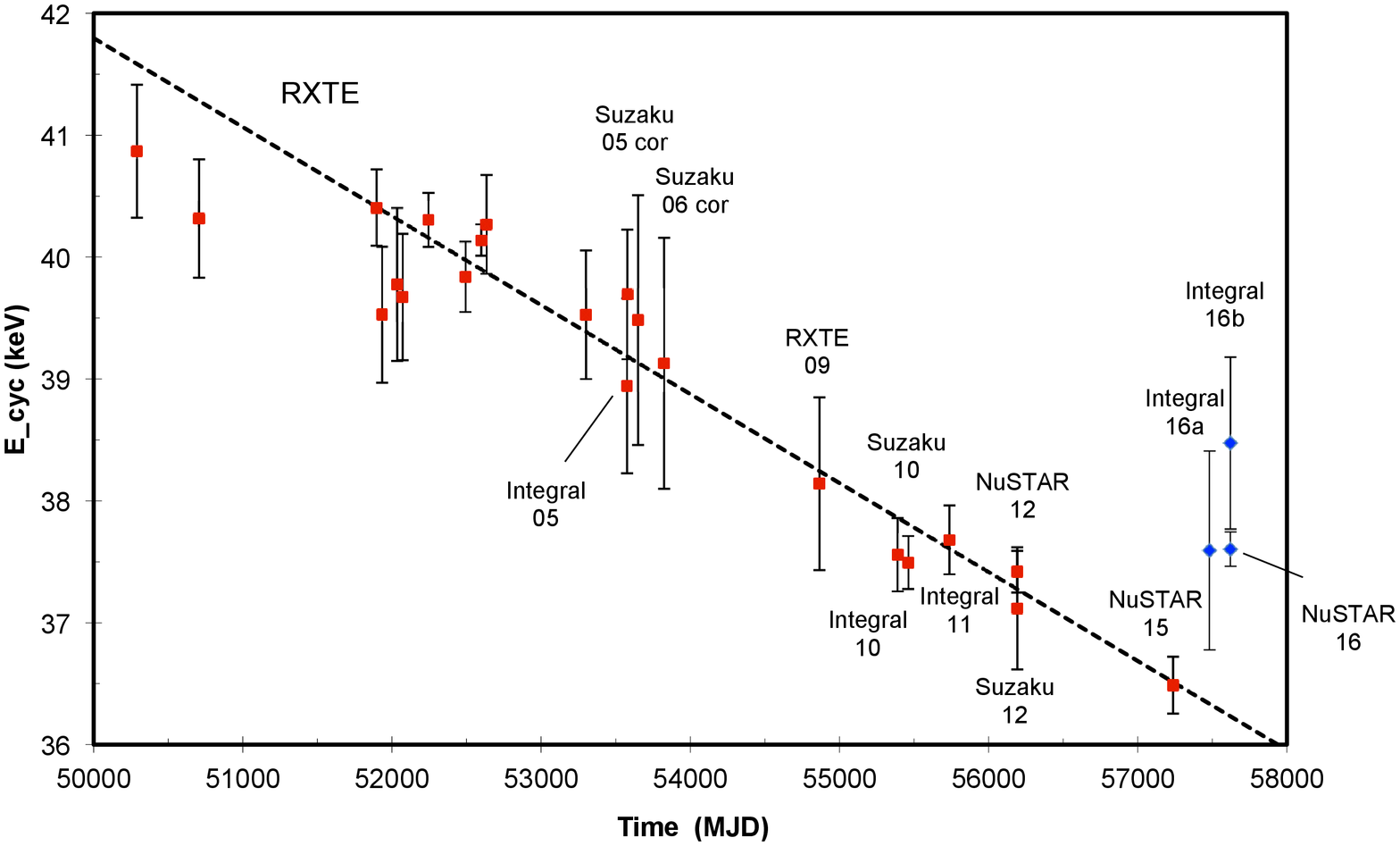}}
\vspace*{-2cm}
\caption{Evolution of the cyclotron line energy $E_\mathrm{cyc}$ in
  Her~X-1. Shown are the pulse phase averaged $E_\mathrm{cyc}$ values
  normalized to a reference ASM count rate of 6.8\,$\mathrm cts\,s^{-1}$ 
  using a flux dependence of 0.44\,keV/$\mathrm ASM-cts\,s^{-1}$. The red 
  points for 1996--2015 (MJD 50000--57300) and the corresponding linear  
  best fit (dashed line) are reproduced from \citet{Staubert_etal16}. The blue 
  points are the new measurements (see Table~\ref{tab:obs}). The flux-normalized 
  values (blue)  are \textsl{NuSTAR} 2016: $37.6\pm0.14$\,keV, \textsl{INTEGRAL} 
  2016a: $37.59\pm0.82$\,keV, and \textsl{INTEGRAL} 2016b: 
  $38.47\pm0.71$\,keV. The 20-year decay of $E_\mathrm{cyc}$ has ended
  with a strong indication of an inversion. }
\end{figure*}

The centroid energy $E_\mathrm{cyc}$ of the CRSF in Her~X-1 is
variable with respect to several parameters, namely pulse phase, luminosity, 
time, and possibly a 35\,d phase \citep{Staubert_etal14}. Apart from the 
discovery of the first CRSF ever, Her~X-1 (through repeated observations) 
has allowed the first discovery of two additional features with respect to its
CRSF, both of which have in the meantime been observed in other binary
X-ray pulsars. First, a positive correlation between $E_\mathrm{cyc}$ and 
the X-ray luminosity $L_{x}$ \citep{Staubert_etal07} (confirmed on short 
timescales by the pulse-amplitude-resolved technique by \citealt{Klochkov_etal11}), 
and second, a long-term decay of $E_\mathrm{cyc}$, co-existing with the 
luminosity dependence \citep{Staubert_etal14,Staubert_etal16}. The long-term 
decay was confirmed by \citet{Klochkov_etal15} using monitoring data of
\textsl{Swift}/BAT (even without taking the luminosity dependence into account).

In addition to  Her~X-1, the following four objects are known to show such a 
positive correlation: A0535+26 \citep{Klochkov_etal11,Sartore_etal15}, Vela~X-1
\citep{Fuerst_etal14b}, GX~304$-$1 \citep{Rothschild_etal16}, and Cep~X-4 
\citep{Vybornov_etal17}. A possible additional source is V0332+53 at low 
luminosity towards the end of an outburst \citep{Doroshenko_etal16}.
At very high luminosity, this object is a good example of a
negative $E_\mathrm{cyc}$--$L$ correlation \citep{Mowlavi_etal06,
Tsygankov_etal10,Cusumano_etal16, Doroshenko_etal16}. With regard
to the long-term variation in $E_\mathrm{cyc}$, there are now three
additional candidates: a decay in Vela~X-1 \citep{LaParola_etal16}, a
decay in V0332+53 during an outburst \citep{Cusumano_etal16,
  Doroshenko_etal16}, and a possible increase in 4U~1538$-$522
\citep{Hemphill_etal16}.

In this {Letter} we present the results of new observations of
Her~X-1 in 2016 of the energy of the cyclotron resonance scattering
feature in the pulse averaged X-ray spectrum of Her~X-1: the CRSF
energy has apparently stopped its 20-year decay and has started
to increase again.

\section{Observations}

In 2016 Her~X-1 was observed by \textsl{INTEGRAL} and \textsl{NuSTAR},
as summarized in Table~\ref{tab:obs}. All observations were close to
the maximum flux of a \textsl{Main-On} state (at
35\,d phases $<0.18$). The observations around 2016 August 21 were
coordinated between \textsl{NuSTAR} and \textsl{INTEGRAL} and are
partly simultaneous. The exposure time of \textsl{NuSTAR} was 37\,ks.
The details of the data analysis are similar to those described by
\citet{Staubert_etal14} and \citet{Staubert_etal16}. We used the
standard \texttt{nupipeline} and \texttt{nuproducts} utilities and
XSPEC\footnote{https://heasarc.gsfc.nasa.gov/xanadu/xspec/} v12.9 as
part of HEASOFT\footnote{http://heasarc.nasa.gov/lheasoft/} 6.18. The
\textsl{INTEGRAL} observations of 100\,ks each were performed in
Revolutions 1663 and 1715. Using the current version of the
\textsl{Off-line Science Analysis} (OSA)
10.2\footnote{http://www.isdc.unige.ch/integral/analysis}, we were not
able to arrive at reliable spectral results with the data from
\textsl{IBIS} because of very large uncertainties in the response and
energy calibration.\footnote{The hope is that the \textsl{IBIS} data
  can be used later, after the release of version 11 of the OSA
  (expected at the end of 2017).} The two data points shown in
Fig.~\ref{fig:Ecyc} are therefore from \textsl{SPI} (see also
\citealt{Ferrigno_etal16}), screened and reduced according to
procedures described by \citet{Churazov_etal11}.

\section{Results and discussion}

Figure~\ref{fig:Ecyc} displays the evolution of the CRSF centroid
energy of Her X-1 from 1996 to 2016, including the data points (in
blue) of the new observations. All other data points (in red) and the
 linear best fit  to the data of 1996--2015 (dashed line) are reproduced from
Fig.~2 of \citet{Staubert_etal16}. We point out that in this
plot all data points are flux corrected. This  is necessary because 
$E_\mathrm{cyc}$ is dependent on time and on flux (or luminosity). 
\citet{Staubert_etal16} have described in detail how the two
dependencies can be separated by a simultaneous fit (in time and
flux). The new data points are also flux corrected by using the same
dependence as before. There is no reason to assume that the flux
dependence has changed (all existing data are reasonably consistent
with one dependence, as demonstrated in Fig.~3 of \citealt{Staubert_etal16}).
We note, however, that even if this dependence had changed somewhat,
the resulting uncertainty would be negligible, since the flux correction
for the new data is very small: from a count rate of 
6.09\,ASM-cts\,$\mathrm{s}^{-1}$ (see Table~\ref{tab:obs}) to the reference 
count rate of 6.80. 
 
It is believed that the variation in $E_\mathrm{cyc}$ with pulse phase
is due to the changing viewing angle under which the emission regions
are seen \citep{Schoenherr_etal07}. The dependence on luminosity (on
both long and short timescales) could be due to a change in the
height of the emission region above the neutron star surface. Within the 
concept of different ``accretion regimes'' \citep[e.g.,][]{Becker_etal12}, the
negative $E_\mathrm{cyc}$ / $L_\mathrm{X}$ correlation in
supercritical accretion ($L_\mathrm{X}\gtrsim
\sim10^{37}\,\mathrm{erg}\,\mathrm{s}^{-1}$, as in, e.g., V0332+53) is
due to the movement of a radiation dominated shock (and the primary
emission region) to larger distances from the neutron star surface  (i.e., to 
a weaker $B$-field), when the accretion rate increases. The opposite is
expected to happen at subcritical accretion rates.
\citet{Staubert_etal07} have shown that a positive correlation is
actually expected in the case of lower accretion rates. Under this
condition, the dynamical pressure of the in-falling material reduces
the height of the emission region when the accretion rate increases,
leading to an increase in $E_{cyc}$,  as was also found by
\citet{Becker_etal12}. Recently, a model involving a collisonless
shock was developed that also explains the deviation from a pure
linear dependence (a ``roll-off''), as observed in GX~304$-$1
\citep{Rothschild_etal17,Vybornov_etal17}.

From Fig.~\ref{fig:Ecyc} it is apparent that the 20-year decay
of $E_\mathrm{cyc}$ does not continue after 2015; instead, an inversion
has happened. By comparing only the last two  (flux corrected) data
points from \textsl{NuSTAR} of 2015 and 2016 the significance of the
turn-up is found to be $\sim$4 standard deviations. If the 2016
\textsl{NuSTAR} point is compared to the value expected at this time
for a continued linear decay, the separation of the observed value
from the dashed line is significant to $\sim$8 standard deviations.
Using the last three \textsl{NuSTAR} points, the decrease of
$(-8.5\pm0.3)\times 10^{-4}\,\mathrm{keV}\,\mathrm{d}^{-1}$ that was
seen between 2012 and 2015 has changed to an increase of
$(35.6\pm7.0)\times 10^{-4}\,\mathrm{keV}\,\mathrm{d}^{-1}$ between
2015 and 2016. We note that the rate of increase is three to five times
faster than that of the last decrease. Even though the uncertainties
of the two \textsl{INTEGRAL/SPI} data points are significantly larger,
we still consider that these points  support a real turn-up.
In addition,  the fact that the 2016 data point from 
\textsl{NuSTAR}  does not fit into the previous picture is evident if 
it is time corrected (using the same time dependence as before);
plotted into Fig.~3 of \citet{Staubert_etal16}, it would sit at 
6.09\,ASM-cts\,$\mathrm{s}^{-1}$ and $E_\mathrm{cyc}$ =
$40.27\pm0.15$\,keV, which is $\sim8.5$ $\sigma$ above the best 
fit central line in this plot.

Because a further increase in $E_\mathrm{cyc}$ might be expected, 
it seems important to observe Her~X-1 over again at short intervals.
The new increase could be similar to that observed during the period
1990--1993 \citep[see][]{Gruber_etal01,Staubert_etal14}, which is
rather fast. Even though the earlier increase was not well covered by
observations, this jump of $\sim7$\,keV over $\sim4$\,years suggests a
rate around ${\sim}5\times 10^{-3}\,\mathrm{keV}\,\mathrm{d}^{-1}$,
which is very similar to the rate found for the onset of the current
increase.

The long-term decay of $E_\mathrm{cyc}$ was securely established when
a model for the simultaneous dependence on luminosity and on time was
applied to the existing data of 1996--2015 \citep{Staubert_etal16}.
The overall drop over 20\,years amounts to $\sim$5\,keV, or
13\%.\footnote{The slight flux reduction of $\sim$15\% between 1996
  and 2016 \citep{Staubert_etal16} contributes only $\sim$0.4\,keV
  when the derived flux dependence of
  $-0.44\,\mathrm{keV}/\mathrm{(ASM-cts/s)}$ is applied. } The lowest
value reached recently is $\sim$37\,keV, which is similar to the value
measured at the time of the discovery of the line in 1975 and for
several years thereafter. This may not be a chance coincidence, but
could have a physical meaning. Under the assumption of a continued decay, 
\citet{Staubert_14} speculated that  a new
turn-up might be expected once $E_\mathrm{cyc}$  reached a
``bottom'' value comparable to the one before the 1990--1993 turn-up.
It seems that this speculation has now come true. It was further
speculated that the observed behavior -- a rather fast rise followed
by a decay over tens of years -- might possibly be a cyclic behavior
that could be understood in the framework of a physical scenario, the
basics of which were outlined by \citet{Staubert_etal14},
\citet{Staubert_14}, and \citet{Staubert_etal16}, and are partly
repeated here.

With regard to the long-term decay of $E_\mathrm{cyc}$ we think that
it is either a geometric displacement of the emission region (away
from the neutron star surface) or a change in the local field
configuration which evolves due to continued accretion, rather than a
change in the strength of the underlying global dipole.
\footnote{Here we do not consider {screening} or {burial} of the
magnetic field by continued accretion, or more exotic effects like
 {Ohmic dissipation} or {hydrodynamic flows}
(see \citealt{Staubert_etal14,Staubert_14}).}  

The whole issue of accretion onto highly magnetized neutron stars in
binary X-ray sources is very complex. There are several fundamental questions:  
What happens to the material that is continuously
accreted? Can material be accumulated in the accretion mound, confined
by the $B$-field? If so, how much, and what effect does this have on
the field? Or is the material somehow lost at the bottom of the mound
 either by leaking to larger areas of the neutron star surface or by
incorporation into the neutron star crust? Is the gain and
loss of material in equilibrium?

We suggest that the long-term decay of $E_\mathrm{cyc}$ may be
connected to a slight imbalance between gain and loss, such that the
structure of the column/mound changes. With an accretion rate of
${\sim}10^{17}\,\mathrm{g}\,\mathrm{s}^{-1}$, a variation on
relatively short timescales does not seem implausible. If the
observed decrease in  $E_{\rm cyc}$ were due to a simple movement of
the resonant scattering region to a greater distance from the neutron
star surface (possibly caused by a slightly larger gain than
loss), the observed $\sim$5\,keV reduction in $E_{\rm cyc}$ over 20
years (0.25\,keV per year) would (for a dipole field) correspond to an
increase in height of $\sim$400\,m (e.g., from 10\,m to 410\,m). The
size of the mound and the magnetic field strength and structure within
the mound could have changed with increasing mass. In modeling
magnetic accretion mounds, \cite{MukhBhatt_12} have shown that an
accumulated mass of $\sim 10^{-12}\,M_{\odot}$, which is accreted
within a few hours, can appreciably change the mound size and the
maximum magnetic field strength as well as the field configuration:
the accreted material could drag out the central field lines radially,
thereby diluting the effective field strength in the center (while
enhancing it at larger radii).

So, what is the reason for the turn-up? One might expect that the
accumulation of mass in the accretion mound can find a natural end,
e.g., when the gas pressure in the accretion mound becomes too large
for the magnetic field. Could it be that the fast rise in $E_{\rm
  cyc}$ represents a special event in which the magnetic field in the
accretion mound rearranges itself as a result of a sudden radial
outflow of substantial amounts of material to larger areas of the
neutron star surface?

In the above scenario the long-term decay of $E_{\rm cyc}$ would
correspond to a phase of continuous buildup  of the accretion mound in
terms of mass (and height?) which is followed by an event of rather
sudden mass loss, associated with a rearrangement of the magnetic
field (and of  mass and height of the mound) to the unperturbed
configuration. This is possibly observed through the fast increase in
$E_{\rm cyc}$. If this scenario is viable, and indeed  cyclic
in nature, we may have now seen one full cycle and measured its length to
$\sim$23 to $\sim$25\,yrs. The degree of the postulated imbalance
between gain and loss of accreted material may be roughly estimated to
${\sim}10^{-5}$ by assuming that the total mass in the accretion mound
of, say $10^{-12}\,M_{\odot}$, is doubled in $\sim$20\,yrs.

One might expect that a dramatic event like a substantial and
relatively fast outflow (over a few years)  from the accretion mound
should show up in other observables. In searching for correlations in
time between the two observed inversions ({turn-ups}) of
$E_{\rm cyc}$ (1990--1993 and 2015--2016) and all the other
observables in Her~X-1 that are variable on longer timescales (many of which
to tens of years), such as luminosity, pulse period, 35\,d turn-on
history, (quasi-)period of the Anomalous Lows (AL), we have found the
following near coincidences (see Table~\ref{tab:AL}): \textsl{turn-up
  1}: 1990--1993 / \textsl{AL 2}: start in Aug.\ 1993; \textsl{turn-up
  2}: Aug.\ 2015--Apr.\ 2016 / \textsl{AL 5}: start in Nov.\ 2015.
Both turn-ups in $E_{\rm cyc}$ are close in time to rather prominent
periods of pulse period increase (spin-down).\footnote{Measured pulse
  periods of Her~X-1 can be found at:
http://legacy.gsfc.nasa.gov/compton/data/batse/pulsar/onboard\_folded/
herx-1 and https://gammaray.nsstc.nasa.gov/gbm/science/pulsars.html}
The separation is $\sim$23\,yr, which is close to four times our estimate
of the mean period ($\sim$5.5\,yrs) with which \textsl{Anomalous Lows}
tend to appear (we note that during these 23\,years one AL did actually
not occur. The (near) coincidences between turn-ups, ALs, and
stretches of significant spin-down may just be by chance, but could
also have a physical meaning. Speculating about this is beyond the
scope of this contribution.

We note that Her~X-1 is the only highly magnetized accreting pulsar
for which repeated observations over longer periods of time exist.
This provided the base for the discovery and further study of the
luminosity and the time dependence of the CRSF, including the new
turn-up. We therefore urge that the source continues to be observed
regularly on short intervals (e.g., every half year). For 2017 and 2018
this seems to be secured through accepted observing proposals for
\textsl{NuSTAR}, \textsl{INTEGRAL}, \textsl{HXMT}, and
\textsl{Astrosat}). At the same time, it would be very important
that theoretical models be developed further.

\begin{table}
\caption[]{Information about all observed  (col.~1) \textsl{Anomalous Lows} in Her~X-1.
The \textsl{Turn-On} times given are for the last observed \textsl{Turn-On} before the AL,
and the first \textsl{Turn-On} after the end of the AL.
Note that with the numbering of col.~2, the mean repetition period of the ALs is 
$\sim5.5$\,yrs (assuming that nos. 2 and 6 were not observed).
The 35\,d cycle numbering is according to \citet{Staubert_etal83}.}
\vspace{-0.5cm}
\begin{center}
\begin{tabular}{llllllll}
\hline\noalign{\smallskip}
Obs   & No.   &  No. of & Turn-     & No. of & Turn-    & Dur.  & Ref.      \\
erved & of     & 35\,d   & On          & 35\,d  & On        &  of     &              \\
ALs   & AL     & cycl.    &                & cycl.   &             &  AL    &              \\
         &          & before &                & after   &             &          &              \\
         &          & AL       & [MJD]      & AL      & [MJD]  & [d]      &              \\
\hline\noalign{\smallskip}
AL 1   & 1       & 119     & 45469     & 127    & 45752    & 283   & 1,2       \\
AL 2   & 3       & 226     & 49205     & 229    & 49308    & 103   & 3          \\
AL 3   & 4       & 284     & 51224     & 302    & 51826    & 602   & 4,5,6,7 \\
AL 4   & 5       & 334     & 52945     & 340    & 53157    & 212   & 8          \\
AL 5   & 7       & 459     & 57302     & 461    & 57372    & 70     & 8          \\
\noalign{\smallskip}\hline
\end{tabular}\end{center}
\vspace{-0.2cm}
References: 1. \textsl{EXOSAT}: \citet{Parmar_etal85}, 2. optical: \citet{Delgado_etal83}, 
3. \textsl{Rosat / ASCA / BATSE / EUVE}:  \citet{Vrtilek_etal94}, 4. \textsl{RXTE}/ASM: 
\citet{LevineCorbet_99}, 5. \textsl{Beppo}SAX: \citet{Parmar_etal99}, 6. \textsl{RXTE}: 
\citet{Coburn_etal00}, 7. \textsl{RXTE}: \citet{Still_etal01a}, 
8. \textsl{RXTE}/ASM / \textsl{Swift}/BAT: R. Staubert, this work. 
\label{tab:AL}
\end{table}

\vspace{2mm}
\begin{acknowledgements}
The motivation for this paper are new observational data taken by the
NASA satellite \textsl{NuSTAR} and the ESA satellite
\textsl{INTEGRAL}. We would like to acknowledge the dedication of all
the people who have contributed to the great success of these
missions, especially the ``schedulers'' for their efforts with respect
to the nonstandard scheduling of the observations of Her~X-1. Earlier
important data were provided by the equally successful NASA missions
\textsl{RXTE} and \textsl{Swift}.
We thank the anonymous referee for useful suggestions.
\end{acknowledgements}


\bibliographystyle{aa}
\bibliography{refs_herx1_Aug17}

\end{document}